\documentclass[twocolumn,pre]{revtex4}
\usepackage{amsmath,graphicx,quoting}

\begin{document}

\title{Anomalous Dimension in a Two-Species Reaction-Diffusion
System}

\author{Benjamin Vollmayr-Lee$^1$, Jack Hanson$^2$, R. Scott
  McIsaac$^3$, and Joshua D. Hellerick$^1$}

\affiliation{$^1$Department of Physics \&\ Astronomy, Bucknell University,
  Lewisburg, PA 17837, USA}

\affiliation{$^2$Department of Mathematics, City College of New York, 160
  Convent Ave, New York, NY 10031 USA}

\affiliation{$^3$Calico Life Sciences, South San Francisco, CA 94080 USA}

\begin{abstract}
  We study a two-species reaction-diffusion system with the reactions
  $A+A\to (0, A)$ and $A+B\to A$, with general diffusion constants
  $D_A$ and $D_B$.  Previous studies showed that for dimensions $d\leq
  2$ the $B$ particle density decays with a nontrivial, universal
  exponent that includes an anomalous dimension resulting from field
  renormalization.  We demonstrate via renormalization group methods
  that the scaled $B$ particle correlation function has a distinct
  anomalous dimension resulting in the asymptotic scaling $\tilde
  C_{BB}(r,t) \sim t^{\phi}f(r/\sqrt{t})$, where the exponent $\phi$
  results from the renormalization of the \textit{square} of the field
  associated with the $B$ particles.  We compute this exponent to
  first order in $\epsilon=2-d$, a calculation that involves 61
  Feynman diagrams, and also determine the logarithmic corrections at
  the upper critical dimension $d=2$.  Finally, we determine the
  exponent $\phi$ numerically utilizing a mapping to a four-walker
  problem for the special case of $A$ particle coalescence in one
  spatial dimension.
\end{abstract}

\maketitle

\section{Introduction}

Reaction-diffusion systems are known to exhibit a strong dependence on
fluctuations in lower dimensions that renders mean-field rate
equations invalid \cite{Ovchinnikov78,Toussaint83}.  For example, the
$A+A\to 0$ annihilation reaction has rate equation $\partial_t \langle
a\rangle=-\Gamma \langle a\rangle^2$, which provides a density
decaying as $\langle a\rangle\sim 1/(\Gamma t)$ with the nonuniversal
rate constant $\Gamma$.  But for dimensions $d\leq d_c=2$ nontrivial
correlations develop and give rise to universal power-law behavior
\cite{Peliti86}, the density decaying as $\langle a\rangle \sim
A(Dt)^{-d/2}$ (with logarithmic corrections in $d=2$), where $D$ is
the diffusion constant and $A$ is a universal amplitude \cite{Lee94}.
Field-theoretic renormalization group (RG) methods have proved useful
in analyzing this fluctuation-dominated regime.  See
Ref.~\cite{Taeuber05} for a review.

For many irreversible reactions, such as the annihilation reaction
above, the resulting dynamical exponents (but not the amplitudes) may
simply be determined from Smoluchowski theory
\cite{Smoluchowski17,Chandrasekhar43}, which is an improved rate
equation with a time-dependent reaction rate.  The success of the
Smoluchowski theory, in spite of being an uncontrolled approximation,
stems from the lack of field renormalization in these theories, and
thus the absence of an anomalous dimension.  Counterexamples are
processes with competing branching reactions, such as
branching-annihilating random walks \cite{Cardy96,Cardy98} or directed
percolation \cite{Hinrichsen00,Janssen05,Taeuber14}, for which field
renormalization is required and nontrivial scaling exponents result.

Nevertheless, nontrivial exponents can arise in reaction-diffusion
systems without branching reactions.  We consider such a model here: two
particle species, $A$ and $B$, with diffusion
constants $D_A$ and $D_B$ undergo the reactions
\begin{align}
  A+A&\to \begin{cases} A & \text{(coalescence) prob. } p \\
    0 & \text{(annihilation) prob. } 1-p.
    \end{cases}  \nonumber\\
  A+B&\to A \quad\text{(trapping).}
  \label{eq:reaction}
\end{align}
The rate equations, valid for $d>2$, are
\begin{equation}
  \partial_t \langle a\rangle = - \Gamma\langle a\rangle^2 \qquad
  \partial_t \langle b\rangle = -  \Gamma' \langle a\rangle \langle b\rangle,
\end{equation}
where the angle brackets represent averages with respect to the 
stochastic processes of diffusion and reaction, as well as over
initial conditions.  These result
in $\langle a\rangle\sim 1/(\Gamma t)$ and $\langle
b\rangle\sim t^{-\theta}$ with the exponent $\theta=\Gamma'/\Gamma$
given by nonuniversal rate constants. The fluctuation dominated
case of $d\leq 2$ has been studied by the Smoluchowski approach
\cite{Krapivsky94b} and by RG techniques
\cite{Howard96,Krishnamurthy03,Rajesh04} (with $A$ particle dynamics
reducing to the well-studied single-species reaction
\cite{Peliti86,Lee94}.)  In contrast to the rate equation result,
$\theta$ is found to be universal, depending only on the parameters
$\delta = D_B/D_A$ and $p$.  Smoluchowski theory gives
\begin{equation}
  \theta_S = \frac{d}{2-p}\biggl(\frac{1+\delta}{2}\biggr)^{d/2},
\end{equation}
while the RG predicts
\begin{equation}
  \theta = \theta_S + \frac{1}{2}\gamma_b^*
  \label{eq:theta}
\end{equation}
where the anomalous dimension $\gamma_b^*$, given in
Eq.~(\ref{eq:gammabstar}), is of order $\epsilon=2-d$ and stems from
the necessary field renormalization of the $b$ density.

In the present work we demonstrate the existence of an
\textit{additional} anomalous dimension $\phi$ for this system which
emerges from the scaled $B$ particle correlation function
\begin{equation}
  \tilde C_{BB}(r,t)
  \equiv \frac{\langle b(r,t) b(0,t) \rangle - \langle b(t)\rangle^2}
         {\langle b(t)\rangle^2} \sim t^\phi f(r/\sqrt{t}).
         \label{eq:CBBscaling}
\end{equation}
In contrast, the scaled correlation functions $\tilde C_{AA}$ and
$\tilde C_{AB}$ are simply functions of $r/\sqrt{t}$ with no time
dependent prefactor.  This exponent results from the multiplicative
renormalization factor $Z_{b^2}$ required by the $b^2$ density
operator, causing $\langle b^2\rangle/\langle b\rangle^2$ to be
renormalized by a factor $Z_{b^2}/(Z_b)^2$, while $\langle
ab\rangle/(\langle a\rangle\langle b\rangle)$ has no corresponding
factor.  The case of $\langle a^2\rangle/\langle a\rangle^2$ is more
subtle: as shown in \cite{Munasinghe06}, a multiplicative
renormalization factor $Z_{a^2}$ is required due to the
anticorrelation of the $A$ particles, but with the consequence that
$\tilde C_{AA}\propto r/\sqrt{t}$ rather than anomalous time
dependence of the form (\ref{eq:CBBscaling}). 

Smoluchowski theory gives $\phi_S=0$ due to the lack of field
renormalization.  We turn to the field-theoretic RG treatment to
obtain a systematic expansion in powers of $\epsilon=2-d$, where the
number of loops in the Feynman diagram expansion is equivalent to the
resulting order of $\epsilon$.  The tree level (zero loop) RG
calculation also gives $\phi = 0+O(\epsilon)$.  At one-loop order the
calculation involves 61 diagrams, giving
\begin{equation}
  \phi = \frac{13}{24-18p}\epsilon + O(\epsilon^2),
  \label{eq:phi}
\end{equation}
which is the primary result of this paper.  The calculation
demonstrates that the exponent is universal, depending only on $p$ and
the diffusion constant ratio $\delta$, though curiously the $\delta$
dependence drops out to first order in $\epsilon$.  Since $\phi$ is
positive, the $B$ density fluctuations increase with time.

As a consequence of the anomalous dimension, the amplitudes of both
$\langle b(t)\rangle$ and $\tilde C_{BB}(0,t)$ are nonuniversal for
$d<2$.  Asymptotically, these quantities are functions of the
universal lengths $\sqrt{D_At}$ and $\sqrt{D_Bt}$ as well as a
nonuniversal length scale $\Lambda$ that reflects the lattice spacing
or capture radius, reaction rates, and other microscopic details.
From dimensional analysis, $\langle b(t)\rangle \sim
t^{-d/2}g(\Lambda/\sqrt{t})\sim \Lambda^{2\theta-d}t^{-\theta}$, and
similarly, $\tilde C_{BB}(0,t) = g(\Lambda/\sqrt{t}) \sim
\Lambda^{-2\phi} t^\phi$.  


At the upper critical dimension $d_c=2$ we obtain the density
\begin{equation}
  \langle b\rangle \sim (\ln t)^{\alpha} t^{-(1+\delta)/(2-p)}.
  \label{eq:density_at_dc}
\end{equation}
where
\begin{equation}
  \alpha = \frac{3}{2}\biggl(\frac{1+\delta}{2-p}\biggr) + \frac{1}{2}\biggl(
  \frac{1+\delta}{2-p}\biggr)^2 f(\delta),
  \label{eq:alpha}
\end{equation}
with $f(\delta)$ given in Eq.~(\ref{eq:f}).  The density decay
exponent is necessarily discontinuous at the upper critical dimension
$d_c=2$, since it is universal below and nonuniversal above $d_c$.
The power law in (\ref{eq:density_at_dc}) corresponds to the
$\epsilon\to 0$ limit of (\ref{eq:theta}), as was found by
\cite{Howard96} and \cite{Rajesh04}; however, our exponent $\alpha$
differs from those previous works.   We return to this point
in the summary.
  We find the scaled correlations at
the upper critical dimension to have the form
\begin{equation}
  \tilde C_{BB}(r,t)\sim (\ln t)^{\alpha_2} f(r/\sqrt t)
  \label{eq:correlations_at_dc}
\end{equation}
with
\begin{equation}
  \alpha_2 = \frac{1+9p}{12-9p}.
  \label{eq:beta}
\end{equation}

For the case where the $A$ particles undergo the coalescence reaction
($p=1$) in one spatial dimension, the exponent $\phi$ can be related
to a four walker problem which we call the \textit{bracket} problem:
given four random walkers that begin spaced along a line, the
probability that the middle two walkers ($B$'s) have not met either of
the end walkers ($A$'s) by time $t$ decays as $t^{-\beta}$.  We
measure this exponent numerically with use of a mapping to an
electrostatic problem \cite{Redner99,benAvraham03}, and find for the
case of equal diffusion constants that $\beta = 1.873754(3)$. Through
scaling arguments presented below, we have (for $p=\delta=d=1$) $\phi
= 5/2 - \beta$, giving
\begin{equation}
  \phi \simeq 0.626246(3).
\end{equation}
In comparison, the $\epsilon$ expansion truncated at first order and
evaluated at $\epsilon=1$ provides $\phi=13/6\simeq 2.17$.  Evidently the
$\epsilon$ expansion is not rapidly convergent.

The layout of the paper is as follows.  In Sec.~\ref{sec:fieldtheory}
we introduce the field theory and diagrammatic expansion.  In
Sec.~\ref{sec:renormalization} we discuss the renormalization of the
theory and review how it applies to the $B$ density.  This is followed
in Sec.~\ref{sec:phicalculation} by a calculation of correlation
function anomalous dimension $\phi$ to order $\epsilon$, with details
of this calculation provided in an appendix. The density and
correlations at the upper critical dimension $d=2$ is discussed in
Sec.~\ref{sec:twodimensions}.  In Sec.~\ref{sec:numerical} we present
the numerical calculation of $\phi$ that exploits a connection to the
four-walker bracket problem.  Finally, we summarize our results in
Sec.~\ref{sec:summary}.

\section{Field Theory}

\label{sec:fieldtheory}

\begin{figure}
\includegraphics[width=2.6in]{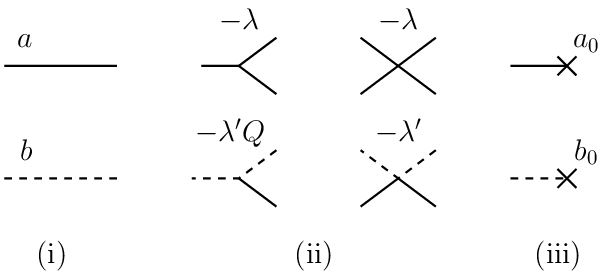}
\caption{Diagram ingredients: (i) the diffusion propagators, (ii) the
  interaction vertices, and (iii) the initial terms of the theory.
  Time flows from right to left.}
\label{fig:ingredients}
\end{figure}

The two-species model described in Eq.~(\ref{eq:reaction}) is first
written in terms of a probability master equation, and then following
standard methods \cite{Doi76,Grassberger80,Peliti85,Taeuber05}
converted via Fock space to a field theory.  The resulting
action is
\begin{align}
  S =
  \int d^dx \, & dt \biggl\{ \bar a(\partial_t - \nabla^2)a
  + \bar b(\partial_t-\delta\nabla^2)b \nonumber\\
  &+\lambda \bar a a^2 + \lambda \bar a^2 a^2 
  +\lambda' Q\bar bab + \lambda' \bar a\bar b ab \nonumber\\
  &+ ( \bar a a_0 + \bar b b_0)\delta(t)
  \biggr\}.
  \label{eq:action}
\end{align}
Here $a$ and $b$ are complex fields corresponding to $A$ and $B$
particles, and $\bar a$ and $\bar b$ are auxiliary fields.  The first
line in (\ref{eq:action}) represents the diffusion process, with time
rescaled so that $D_A= 1$ (recall $\delta=D_B/D_A$).  The second line
represents the reaction processes with microscopic rate constants
$\lambda$ for annihilation and coalescence and $\lambda'$ for the
trapping reaction.  For notational simplicity we have introduced the
parameter $Q=1/(2-p)$.  In the conventional mapping, a factor $1/Q$
appears in the $\bar aa^2$ coefficient, reflecting the average number
of particles removed by an $A+A$ reaction.  In the action above we
have rescaled $a\to Qa$, $\bar a\to \bar a/Q$ for convenience.  The
third line corresponds to Poissonian initial conditions with average
densities $a_0$ (after rescaling) and $b_0$.  The components of
Feynman diagrams resulting from this action are shown in
Fig.~\ref{fig:ingredients}.

\begin{figure}
  \includegraphics[width=2.4in]{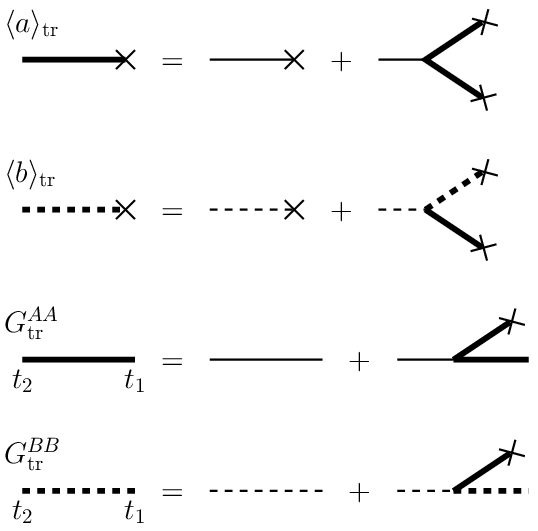}
  \caption{Diagrammatic representation of the Dyson equations for the
    tree level densities and dressed propagators, which are depicted
    as bold lines.}
  \label{fig:dysonsums}
\end{figure}

The averages of the $a$ and $b$ fields can be directly related to the
densities and correlation functions of the $A$ and $B$ particles.
These can be computed as an expansion in the number of loops, which
results after renormalization in an $\epsilon=2-d$ expansion.  An
infinite number of diagrams result for each order in the loop
expansion.  To evaluate these infinite sums we need the tree-level
(zero loop) densities $\langle a\rangle_\text{tr}$ and $\langle
b\rangle_\text{tr}$ and dressed propagators
$G^{AA}_\text{tr}({\bf k},t_2,t_1) =\langle a({\bf k},t_2) \bar
a(-{\bf k},t_1)\rangle_\text{tr}$ and $G^{BB}_\text{tr}({\bf
  k},t_2,t_1)=\langle b({\bf k},t_2) \bar b(-{\bf
  k},t_1)\rangle_\text{tr}$.  These can be computed from Dyson
equations \cite{Lee94,Rajesh04,Taeuber05}, shown in
Fig.~\ref{fig:dysonsums}, resulting in
\begin{align}
  \langle a(t)\rangle_\text{tr} &= \frac{a_0}{1+\lambda a_0 t} \\
  \langle b(t)\rangle_\text{tr} &=
  \frac{b_0}{(1+\lambda a_0 t)^{Q\lambda'/\lambda}} 
\end{align}
and for $t_2>t_1$
\begin{align}
  G^{AA}_\text{tr}({\bf k},t_2,t_1) &=
  \biggl(\frac{1 + a_0 \lambda t_1}{1 + a_0\lambda t_2}\biggr)^2
  e^{-k^2(t_2-t_1)} \\
  G^{BB}_\text{tr}({\bf k},t_2,t_1) &=
  \biggl(\frac{1 + a_0 \lambda t_1}{1 + a_0\lambda t_2}
  \biggr)^{Q\lambda'/\lambda}
  e^{-\delta k^2(t_2-t_1)},
\end{align}
with $G^{AA}_\text{tr} = G^{BB}_\text{tr} = 0$ for $t_2<t_1$.  Notice
that the initial density contributions break time translation
invariance in the dressed propagators.

\begin{figure}
  \includegraphics[width=2.4in]{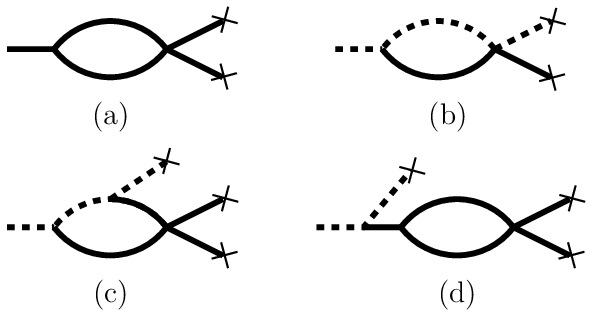}
  \caption{(a) One loop contributions to $\langle a(t)\rangle$ and
    (b)--(d) one loop contributions to $\langle b(t)\rangle$,
    constructed from the tree level densities and dressed
    propagators.}
  \label{fig:oneloopdensities}
\end{figure}

\begin{figure}
  \includegraphics[width=3.2in]{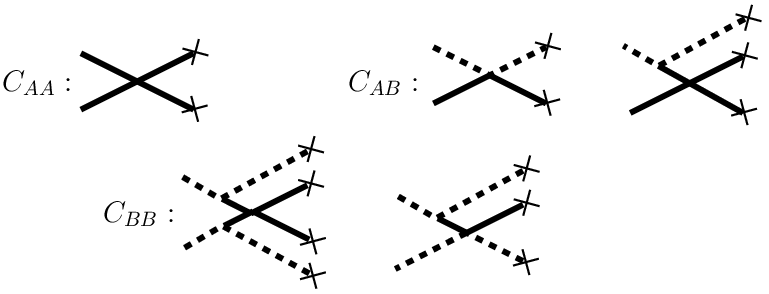}
  \caption{Tree level correlation functions.}
  \label{fig:treelevelcorrelations}
\end{figure}

With these tree level quantities we can calculate, for example, all
one-loop diagrams contributing to $\langle a(t)\rangle$ and $\langle
b(t)\rangle$.  These require one terminal $a$ or $b$ propagator at
time $t$, as shown in the Feynman diagrams of
Fig.~\ref{fig:oneloopdensities}.  Additionally, the tree level
diagrams for the (unscaled) correlation functions $C_{AA}(r,t)=\langle
a(r,t) a(0,t)\rangle-\langle a(t)\rangle^2$ and similarly defined
$C_{AB}$ and $C_{BB}$ are shown in
Fig.~\ref{fig:treelevelcorrelations}.

Note the similar topology of the diagrams.  The $\langle b(t)\rangle$
diagrams in Fig.~\ref{fig:oneloopdensities} can be constructed
from diagram (a) by inserting a dashed $B$ line either before the
loop, inside the loop, or after the loop, resulting in diagrams (b),
(c), and (d) respectively.  Similarly, the $C_{AB}$ and $C_{BB}$
correlations are constructed by adding $B$ lines to the $C_{AA}$
diagram.  This technique proves useful in generating the one-loop
contributions to $C_{BB}$.

\section{Renormalization}

\label{sec:renormalization}

The bare diagrammatic loop expansion fails to converge for large $t$
when $d\leq 2$, indicating the necessity of renormalizing the theory.
Following standard procedure \cite{Taeuber05}
we introduce a normalization time $t_0$ and define the dimensionless
coupling constants $g_0 = \lambda t_0^{\epsilon/2}$ and $g'_0 =
\lambda' t_0^{\epsilon/2}$.  The renormalization of the vertices takes
the unusually simple form of a geometric sum
\cite{Peliti86,Lee95,Howard96}, giving
\begin{equation}
  g_R = \frac{g_0}{1 + g_0/g_*}, \qquad
  g'_R = \frac{g'_0}{1 + g'_0/g'_*}, \qquad
\end{equation}
with
\begin{align}
  g_* &= \frac{(8\pi)^{d/2}}{2\Gamma(\epsilon/2)} = 2\pi \epsilon +
  O(\epsilon^2) \\
  g'_* &= \frac{[4\pi(1+\delta)]^{d/2}}{\Gamma(\epsilon/2)} =
  (1+\delta)2\pi \epsilon + O(\epsilon^2).
\end{align}
For the two-species reaction of  Eq.~(\ref{eq:reaction}), additional field
renormalization is required for $B$ particles: the density $b_B$
calculated from the bare theory is related to the renormalized density
$b_R$ via $b_B = Z_b b_R$, with $Z_b(g_R,g_R')$ chosen to ensure that
the expansion of $b_R$ in powers of $g_R$ has nonsingular coefficients
in $\epsilon$.  From dimensional analysis and the fact that $b_B$ does
not depend on the normalization time $t_0$ we obtain the RG equation
\begin{align}
  \biggl[ t\frac{\partial}{\partial t} - 
    &\frac{d}{2}a_0\frac{\partial}{\partial a_0} +
    \frac{1}{2}\beta(g_R)\frac{\partial}{\partial g_R} +
    \frac{1}{2}\beta(g'_R)\frac{\partial}{\partial g'_R} \nonumber\\  
    &+ \frac{1}{2} \gamma_b(g_R,g_R')\biggr] b_R(t,a_0,g_R,g_R';t_0) = 0,
  \label{eq:rg_equation}
\end{align}
where
\begin{align}
  \beta(g_R) &\equiv -2t_0\frac{\partial}{\partial t_0} g_R
  = -\epsilon g_R + \frac{\epsilon}{g_*} g_R^2, \\
  \beta(g'_R) &\equiv -2t_0\frac{\partial}{\partial t_0} g'_R
  = -\epsilon g'_R + \frac{\epsilon}{g'_*} {g'_R}^2, 
\end{align}
and
\begin{equation}
  \gamma_b(g_R,g_R') \equiv  -2t_0\frac{\partial}{\partial t_0}
  \ln Z_b,
\end{equation}
with $Z_b$ yet to be determined.  Eq.~(\ref{eq:rg_equation}) is solved
by the method of characteristics, leading to the asymptotic solution
\begin{align}
  b_R(&t,a_0,g_R,g_R'; t_0)  \nonumber\\
  & \sim (t/t_0)^{-\gamma_b^*/2}
  b_R\Bigl(t_0, a_0(t/t_0)^{d/2}, g_*, g'_*; t_0\Bigr),
  \label{eq:method_of_characteristics}
\end{align}
where $\gamma_b^*=\gamma_b(g_*, g'_*)$.
The general strategy is to compute the bare Feynman diagrams and
express the couplings in terms of $g_R$ and $g'_R$, which then flow to
their fixed points $g_*$ and $g'_*$ on the right hand side of
Eq.~(\ref{eq:method_of_characteristics}). Then the asymptotic time
dependence of $b_R$ is determined by the renormalized $a_0$ and
the anomalous dimension $\gamma_b^*$.

The bare tree level and one loop diagrams for the $B$ density have
been evaluated in Refs.~\cite{Howard96,Rajesh04} and can be written as
\begin{equation}
  b_B = \frac{b_0}{(1+a_0 \lambda t)^{Q\lambda'/\lambda}}\biggl[
    1 + \lambda t^{\epsilon/2}\biggl(\frac{A(z)}{\epsilon^2} +
    \frac{B(z)}{\epsilon}+\dots\biggr)\biggr]
  \label{eq:bB}
\end{equation}
with the coupling constant ratio 
\begin{equation}
  z\equiv\frac{\lambda'}{\lambda}=\frac{g_0'}{g_0}.
\end{equation}
The ratio of bare coupling constants can be expressed in terms of the
renormalized couplings as
\begin{equation}
  z  = \frac{g_R'}{g_R} \biggl[1 + \sum_{k=0}^\infty
    \biggl(\frac{g_R'}{g_*'}-\frac{g_R}{g_*} \biggr)
    \frac{{g_R'}^k}{{g_*'}^k}\biggr].
\end{equation}
As $g_R$ and $g_R'$ flow to their fixed point values, all terms in the
expansion cancel and so $z$ flows to the fixed point value
\begin{equation}
  z^*=\frac{g'_*}{g_*} = 2\biggl(\frac{1+\delta}{2}\biggr)^{d/2}
  = 1+\delta + O(\epsilon).
\end{equation}
Since $A(z)= -Qz(8\pi)^{\epsilon/2}\pi^{-1}(1-z/z^*)$ vanishes as
$z\to z^*$, the $A(z)/\epsilon^2$ contribution under renormalization
is subleading in time and can be neglected.  Similarly, $B(z)$ can be
expanded in powers of $z-z^*$ with only the leading
\begin{equation}
  B(z^*) = \frac{1}{4\pi}\biggl(3Q(1+\delta)  + Q^2(1+\delta)^2
  f(\delta)\biggr) + O(\epsilon)
\end{equation}
contributing to the asymptotic time dependence, where
\begin{equation}
  f(\delta) = 1 + 2\delta\biggl[\ln\biggl(\frac{2}{1+\delta}\biggr)
    -1\biggr]
  + (1-\delta^2)\biggl[ \text{Li}_2\biggl(\frac{\delta-1}{\delta+1}\biggr)
    -\frac{\pi^2}{6}\biggr]
  \label{eq:f}
\end{equation}
and $\text{Li}_2(v)=-\int_0^v du\, \ln(1-u)/u$ is the dilogarithm
function \cite{AbramowitzStegun}.  A useful special case is $f(1)=-1$.

Substituting $t\to t_0$, $a_0\to a_0(t/t_0)^{d/2}$, and $\lambda\to
t_0^{-\epsilon/2}g_0 = t_0^{-\epsilon/2}(g_R + g_R^2/g_* +\dots)$ into
(\ref{eq:bB}) and then expanding in powers of $g_R$ and $g_R'$ gives
to linear order
\begin{equation}
  b_B = \frac{b_0}{(a_0 g_R t^{d/2})^{Qg_R'/g_R}}\biggl[
    1 +  \frac{B(z^*)}{\epsilon}g_R -\frac{Q}{g_*}g'_R  + \dots
    \biggr].
  \label{eq:density_gR_exp}
\end{equation}
The $1/\epsilon$ coefficient in the $g_R$ expansion is evidence that
field renormalization is required, and allows us to identify $Z_b$
to linear order in the couplings and leading order in $\epsilon$ as
\begin{equation}
  Z_b = 1 +  \frac{B(z^*)}{\epsilon} g_R -\frac{Q}{g_*}g'_R + \dots
\end{equation}
which in turn gives
\begin{equation}
  \gamma_b = -B(z^*) g_R + \frac{Q}{2\pi} g_R'.
\end{equation}
Evaluating this at the fixed point couplings and recalling that
$Q=1/(2-p)$ gives
\begin{equation}
  \gamma_b^* = -\biggl[\frac{1}{2}\biggl(\frac{1+\delta}{2-p}\biggr)
     +\frac{1}{2}
    \biggl(\frac{1+\delta}{2-p}\biggr)^2 f(\delta)\biggr]
   \epsilon + O(\epsilon^2),
  \label{eq:gammabstar}
\end{equation}
with $f(\delta)$ from Eq.~(\ref{eq:f}).
Finally, the renormalized asymptotic density $b_R$ then has via
Eq.~(\ref{eq:method_of_characteristics}) the asymptotic time
dependence
\begin{equation}
  b_R = Z_b^{-1} b_B \sim t^{-dQz^*/2 -
    \gamma_b^*/2} = t^{-\theta}.
\end{equation}
The renormalized tree level contribution $dQz^*/2$ is exactly the
Smoluchowski exponent, thus we obtain Eq.~(\ref{eq:theta}) as claimed.

Our result (\ref{eq:theta}) matches two previous RG calculations which
utilized slightly different techniques: in Ref.~\cite{Howard96} the
$\epsilon\to 0$ and large $t$ limit were swapped and an expansion in
$\ln t$ was resummed, and in Ref.~\cite{Rajesh04} the logarithmic
derivative of the density was renormalized instead of the density.
The field renormalization approach presented here was previously
employed for the special case of $\delta=1$ \cite{Krishnamurthy03},
which we have extended to general $\delta$.  As we shall show in the
next section, this method is useful for determining the correlation
function anomalous dimension $\phi$.

\section{$\phi$ Calculation}

\label{sec:phicalculation}

An additional independent dynamical exponent appears in the $B$
particle correlation function, resulting from the fact that the $b^2$
field requires a distinct renormalization constant $Z_{b^2}$ that is
not equivalent to $(Z_b)^2$.  We demonstrate this by calculating the
(unscaled) correlation function $C_{BB}({\bf r},t)= \langle b({\bf
  r},t)b(0,t)\rangle - \langle b(t)\rangle^2$.  For simplicity, we
work with the Fourier transform
\begin{equation}
  \hat C_{BB}({\bf  k},t) = \int C_{BB}({\bf r},t) e^{i{\bf k}\cdot{\bf r}} \, d^dx
\end{equation}
at ${\bf k}=0$.  Assuming the scaling form (\ref{eq:CBBscaling}), we expect
$\hat C_{BB}(k=0,t)\sim t^{\phi - 2\theta + d/2}$.  We introduce the
renormalization constant $Z_{b^2}$ which relates the bare correlation
function to the renormalized one, $\hat C_{BB}^B = Z_{b^2} \hat C_{BB}^R$,
and obtain a similar RG equation
\begin{align}
  \biggl[ t\frac{\partial}{\partial t} -
    &\frac{d}{2}a_0\frac{\partial}{\partial a_0} +
    \frac{1}{2}\beta(g_R)\frac{\partial}{\partial g_R} +
    \frac{1}{2}\beta(g'_R)\frac{\partial}{\partial g'_R}
    \nonumber\\ &+ \frac{1}{2} \gamma_{b^2}(g_R,g_R')-\frac{d}{2}\biggr] \hat
  C_{BB}^R(t,a_0,g_R,g_R';t_0) = 0,
  \label{eq:rg_equation_ii}
\end{align}
with
\begin{equation}
  \gamma_{b^2}(g_R,g_R') \equiv -2t_0\frac{\partial}{\partial t_0} \ln Z_{b^2}
\end{equation}
yet to be determined.  The method of characteristics solution is
\begin{align}
  \hat C_{BB}^R(&t,a_0,g_R,g_R'; t_0)  \nonumber\\
  &\sim\biggl(\frac{t}{t_0}\biggr)^{d/2-\gamma_{b^2}^*/2}
  \hat C_{BB}^R\Bigl(t_0, a_0(t/t_0)^{d/2}, g_*, g'_*; t_0\Bigr).
  \label{eq:method_of_characteristics_ii}
\end{align}

The tree-level diagrams in Fig.~\ref{fig:treelevelcorrelations} and
the one loop diagrams in the appendix give the bare result
\begin{align}
  \hat C^B_{BB} = \frac{b_0^2 t{\lambda'}^2 h(Q)}
       {\lambda(a_0\lambda t)^{2Q\lambda'/\lambda}}\biggl[
    1 +  \lambda t^{\epsilon/2}\biggl(&\frac{C(z)}{\epsilon^2}
    + \frac{D(z)}{\epsilon} +\dots\biggr)\biggr],
\end{align}
where we have taken the large $a_0$ limit in anticipation of the RG
flow, and $h(Q)=Q(1-2Q/3)$.  The calculation of the one-loop terms
involves 61 distinct Feynman diagrams, the details of which are
presented in the appendix.  From Eqs.~(\ref{eq:class2}) and
(\ref{eq:class4}) we find, as with the density calculation, that
$C(z)\propto z-z*$, making this term subleading in time.  From
Eqs.~(\ref{eq:class1}), (\ref{eq:class2}), (\ref{eq:class3}), and
(\ref{eq:class4}) we obtain
\begin{align}
  D(z^*) =& - \frac{9 - 19Q -27 Q(1+\delta)+ 18Q^2(1+\delta)}{6\pi(3 - 2 Q)} 
  \nonumber\\  &+  \frac{Q^2(1+\delta)^2 f(\delta)}{2\pi} 
  + O(\epsilon).
\end{align}
The procedure described after Eq.~(\ref{eq:density_gR_exp}) provides
$\hat C_{BB}^B$ as an expansion in the renormalized couplings
\begin{align}
  \hat C_{BB}^B = \frac{b_0^2 g_R'^2 h(Q)}
       {g_R(a_0g_R t^{d/2})^{2Qg_R'/g_R}}&
  \biggl[1+ \biggl(\frac{D(z^*)}{\epsilon}-\frac{1}{g_*}\biggr) g_R \nonumber\\
    &+\biggl(\frac{2}{g_*'} - \frac{2Q}{g_*}\biggr)g_R' + \dots
    \biggr]
\end{align}
The singular coefficients of the $g_R$ and $g_R'$ expansions again
indicates the need for field renormalization, so we identify to linear
order in the couplings and to leading order in $\epsilon$
\begin{equation}
  Z_{b^2} = 1 + \biggl(\frac{D(z^*)}{\epsilon}-\frac{1}{g_*}\biggr)g_R
  +\biggl(\frac{2}{g_*'} - \frac{2Q}{g_*}\biggr)g_R' 
\end{equation}
which gives
\begin{equation}
  \gamma_{b^2} = \biggl(-D(z^*)+\frac{1}{2\pi}\biggr) g_R +
  \biggl(\frac{Q}{\pi}-\frac{1}{\pi(1+\delta)}\biggr) g_R'.
\end{equation}
This evaluates at the fixed point couplings to
\begin{equation}
  \gamma_{b^2}^* = -\biggl[\frac{13}{12-9p}  + \frac{1+\delta}{2-p} 
  + \biggl(\frac{1+\delta}{2-p}\biggr)^2 f(\delta)\biggr]\epsilon 
  + O(\epsilon^2).
\end{equation}
We obtain from (\ref{eq:method_of_characteristics_ii}) the asymptotic
form of the renormalized correlation function 
\begin{equation}
  \hat C_{BB}^R(k=0)\sim t^{d/2-\gamma_{b^2}^*/2 -dQz^*}
\end{equation}
Comparison with the expected time dependence of $t^{d/2-2\theta+\phi}$
results in
\begin{equation}
  \phi = \gamma_b^* - \frac{1}{2}\gamma_{b^2}^*
\end{equation}
which evaluates to the expression given in Eq.~(\ref{eq:phi}).

\section{Logarithmic Corrections in $d=2$}

\label{sec:twodimensions}

At the upper critical dimension of $d=2$ the $\beta$-functions
for the renormalized couplings become
\begin{equation}
  \beta(g_R) = \frac{g_R^2}{2\pi} \qquad
  \beta(g_R') = \frac{{g_R'}^2}{2\pi(1+\delta)}.
\end{equation}
These result in the asymptotic running couplings going to zero as
$\tilde g_R\sim 4\pi/\ln t$ and $\tilde g_R'\sim 4\pi(1+\delta)/\ln
t$, typical for marginal operators.  Plugging these into the
$\gamma$-functions gives
\begin{equation}
  \gamma_b(\tilde g_R,\tilde g_R')\sim 2\Gamma_b/\ln t 
  \qquad
  \gamma_{b^2}(\tilde g_R,\tilde g_R')\sim 2\Gamma_{b^2}/\ln t
  \label{eq:gamma_at_dc}
\end{equation}
where
\begin{equation}
  \Gamma_b = \lim_{\epsilon\to 0}\frac{\gamma_b^*}{\epsilon}
  \qquad
  \Gamma_{b^2} =\lim_{\epsilon\to 0}\frac{\gamma_{b^2}^*}{\epsilon}
  \label{eq:Gammas}
\end{equation}

Inserting this into the density RG equation (\ref{eq:rg_equation})
gives the asymptotic solution
\begin{align}
  b_R(&t,a_0,g_R,g_R';t_0) = \nonumber\\
  &\sim \ln(t/t_0)^{-\Gamma_b} b_R
  \Bigl( t_0,a_0t/t_0,\tilde g_R(t),\tilde g_R'(t);
  t_0\Bigr)
\end{align}
which results in the density given in Eqs.~(\ref{eq:density_at_dc})
and (\ref{eq:alpha}).  As a check on this result, for $p=0$ ($A$
particle annihilation) and $\delta=1$ the $B$ particle density should
match that of the $A$ particles, which is known to decay as $\ln t/t$
in $d=2$ \cite{Toussaint83,Peliti86}. Our expression is consistent
with this.

Inserting the running couplings into the RG equation
(\ref{eq:rg_equation_ii}) for the (unscaled) correlations $\hat
C_{BB}(k=0)$ gives the method of characteristics solution
\begin{align}
  &\hat C_{BB}^R(t, a_0, g_R, g_R'; t_0) \nonumber\\ &\quad \sim
  \biggl(\frac{t}{t_0}\biggr)\ln(t/t_0)^{-\Gamma_{b^2}} 
  \hat C_{BB}^R\Bigl(t_0, a_0t/t_0,
  \tilde g_R(t), \tilde g'_R(t); t_0\Bigr).
\end{align}
This results in the asymptotic time dependence
\begin{equation}
  \hat C_{BB}^R(k=0) \sim t^{1-2(1+\delta)/(2-p)} (\ln t)^{-\Gamma_{b^2}-1
    + 2(1+\delta)/(2-p)}.
\end{equation}
Transforming back to real space and dividing by the density squared
results in $\beta = 2\Gamma_b-\Gamma_{b^2}-1$ and the correlation
function scaling given in Eqs.~(\ref{eq:correlations_at_dc}) and
(\ref{eq:beta}).


\section{Numerical Solution for $A+A\to A$ in $d=1$}

\label{sec:numerical}

In one spatial dimension and for the case $p=1$ ($A$ particle
coalescence) it is known that the density decay exponent $\theta$ can
be determined from the problem of three vicious walkers
\cite{Redner01}.  For any $B$ particle there are nearest neighbor $A$
particles to the left and right which undergo simple random walks,
since the product of any future coalescence events can be identified
as the original $A$ neighbor.  Appealing to the universality of
$\theta$, we may consider the limit of an infinite reaction rate for
$A+B\to A$, which implies the $B$ particle density decay is equivalent
to the survival probability of the three vicious walker problem
\cite{Fisher88b}, giving
\begin{equation}
  \theta = \frac{\pi}{2\arccos(\delta/(1+\delta))}.
\end{equation}

In a similar way, the anomalous dimension $\phi$ of the $C_{BB}$
correlation function is related to a four walker problem in which
middle two walkers ($B$'s) are allowed to meet any number of times,
but have had no encounters with the leftmost and rightmost walkers
($A$'s), i.e., the walkers' positions obey $x_1 < x_2, x_3 < x_4$.  We
refer to this as the \textit{bracket} problem.  Further, the power-law
decay of $\langle b^2\rangle\sim t^{-2\theta+\phi}$ is given by the
probability that not only have both middle walkers survived, but
they have also reached the same location ($x_2=x_3$) at time $t$.
This exponent cannot be simply determined analytically, but it is
possible to map the calculation to an electrostatic problem, which
allows for an accurate numerical determination.  In what follows we
limit consideration to the equal diffusion constant case, $\delta=1$.

Since the center of mass motion plays no role in the absorption
probability, the coordinates $x_i$ of the four walkers can be
projected to the three dimensional subspace $x_1+x_2+x_3+x_4=0$ and
parametrized in terms of the coordinates $u_i = (x_i+x_4)/2$ for $i=1,
2, 3$.  The four walker dynamics maps to isotropic diffusion in this
three dimensional space, the geometry of which is clearly mapped out
in Ref.~\cite{benAvraham03}: the six planes $u_k=\pm u_\ell$, $k\neq
\ell$ correspond to the six possible particle encounters $x_i=x_j$,
$i\neq j$ and divide space into 24 wedges, each corresponding to a
permutation of the ordering of the four walkers.  The four vicious
walker problem reduces to the survival probability of a diffusing
particle in the wedge $|u_1|<u_2<u_3$ with absorbing boundary
conditions at $u_1=\pm u_2$ and $u_2=u_3$.  The bracket problem
corresponds instead to a wedge $|u_1|<u_2$ and $|u_1|<u_3$ with
absorbing boundary conditions at $u_1 = \pm u_2$ and $u_1 = \pm u_3$.
By symmetry the bracket problem wedge is equivalent to the smaller
vicious walker wedge with a reflective boundary at $u_2=u_3$.

The time-dependent probability density $p({\bf r},t)$ of a walker that
starts at ${\bf r}_0$ obeys the diffusion equation with
initial condition $p({\bf r},0) = \delta({\bf r}-{\bf r}_0)$ and
$p({\bf r}',t)=0$ for ${\bf r}'$ on the
boundary.  Because of the absorbing boundaries, the probability
density is not normalized for $t>0$.  For such scale-free wedges the
asymptotic behavior for $r\gg r_0$ and $\sqrt{Dt}\gg r_0$ is given by
\begin{equation}
  p({\bf r},t) = C r^\eta t^{-\eta-d/2}
  e^{-r^2/4Dt} f(\hat{\bf r})
  \label{eq:probability}
\end{equation}
where the exponent $\eta$ is related to the smallest eigenvalue of the
spherical laplacian $\nabla^2_{S_d}$ acting in the wedge geometry with
Dirichlet boundary conditions, and $f(\hat{\bf r})$ is the
corresponding eigenfunction \cite{Hammer14,Alfasi15}.  The constant
$C$ depends on the initial location of the particle.

In simple geometries, such as a cone, this eigenvalue problem can be
solved analytically \cite{BenNaim10} to obtain the value of $\eta$,
but this is difficult for  most wedges.  Instead we exploit the fact
that
\begin{equation}
  V({\bf r}) \equiv \int_0^{\infty}  p({\bf r},t)\, dt
  \label{eq:potential}
\end{equation}
obeys Poisson's equation with source $(1/D)\delta({\bf r}-{\bf r}_0)$
\cite{Redner99}.  We can solve this electrostatic problem numerically
to find the large $r$ behavior $V\sim r^{-\mu}$,
and then (\ref{eq:potential}) implies  $\eta = \mu-d+2$.

Finally, the survival probability of the bracket problem is given by
$S(t)=\int p({\bf r}, t)\,d^dr \sim t^{-\beta}$, where the integral is
over the wedge volume. From (\ref{eq:probability}) it follows that
$\beta =\eta/2$.  However, for our exponent $\phi$ we need to impose
the additional constraint that the two center walkers meet at time
$t$, which reduces the dimension of the spatial integral by one, i.e.,
\begin{equation}
  \langle b^2(t)\rangle \sim \bar S(t) \sim 
  \int  p({\bf r}, t)\, d^{d-1}r \sim t^{-\bar\beta},
\end{equation}
giving $\bar\beta = (\eta+1)/2$.  Using the known value of $\theta=3/2$
for this equal diffusion constant case, we get $\phi = 3 -\bar\beta
 = 3-\mu/2$.

We solved Poisson's equation for the bracket problem by
successive over relaxation on an integer lattice with
$u_i^\text{max}=500$.  A point charge was located at $(u_1,u_2,u_3) =
(0,1,1)$, with absorbing boundary conditions at $u_1=\pm u_2$ and
reflecting boundary conditions at $u_2=u_3$.  Following
\cite{benAvraham03}, we did two separate calculations with the
boundary conditions at the edge of the box, $u_3=u_3^\text{max}$,
taken to be either absorbing ($V=0$) or reflecting ($\partial
V/\partial u_3=0$).  The resulting solutions bound the infinite wedge
solution.  We fit the data to the form $V\sim A r^{-\mu}(1 + B
r^{-2})$ to account for the discreteness of the lattice laplacian and
estimate the uncertainty by fitting $V(r)$ along seven different
directions: (0,1,1), (0,1,2), (1,2,2), (0,1,3), (0,2,3), (1,2,3), and
(0,1,4).  We obtain $\mu = 4.747507(6)$, which implies $\eta =
3.747507(6)$ in Eq.~(\ref{eq:probability}) and the values for $\beta$
and $\phi$ reported in the introduction.

\section{Summary and Future Work}

\label{sec:summary}

We have shown that the two-species reaction diffusion system described
by Eq.~(\ref{eq:reaction}) exhibits anomalous dimension, not only in
the $B$ particle density but also in the $BB$ correlation function.
We demonstrated the universality of the anomalous scaling of the
correlation function, Eq.~(\ref{eq:CBBscaling}), and computed the
exponent $\phi$ to first order in $\epsilon$.  Surprisingly the first
order term exhibits no dependence on the diffusion constant ratio
$\delta$.  The exponent $\phi$ goes to zero as $d\to 2$ from below,
continuously connecting to the $d>2$ value, in contrast to the density
decay exponent.  

At the critical dimension $d=2$ we have shown that both the $B$
particle density and the scaled $\tilde C_{BB}$ correlation function
acquire logarithmic corrections, and have computed the associated
exponents.  Our results match previous calculations of the density
decay exponent, \cite{Howard96,Rajesh04}, though differing on the
exponent $\alpha$ for the logarithmic corrections.  As previously
noted \cite{Rajesh04}, Ref.~\cite{Howard96} did not fully incorporate
loop corrections.  Our discrepancy with Ref.~\cite{Rajesh04} is more
troublesome.  They found for $\delta\neq 1$ an additional nonuniversal
contribution to the logarithm exponent that does not arise in our
approach, and supported their calculation with numerical evidence
for $\delta=0$.
Their technique was to renormalize the logarthmic derivative of the
bare density $t\partial_t\ln\langle b(t)\rangle$, rather than
renormalizing the density itself.  It is possible that these two
approaches, which both involve a concurrent $\tilde a_0\to\infty$
limit, are not equivalent.  This issue merits future study.

We derived the anomalous dimension by renormalizing the $b$ and $b^2$
fields by the factors $Z_b$ and $Z_{b^2}$.  Since the density is
exactly proportional to the initial density $b_0$ and the correlation
function is exactly proportional to $b_0^2$, one may equivalently view
the constants $Z_b$ and $Z_{b^2}$ as renormalizing $b_0$ and $b_0^2$.
This has an appealing physical interpretation: as the $A$ particle
correlations approach their universal scaling form, the effective
reaction rate is renormalized, requiring an adjustment in the initial
density of $B$ particles to compensate.  That $b_0^2$ requires a
distinct renormalization indicates that the \textit{fluctuations} in
the $B$ particles must be modified as well.  An interesting direction
for future work would be to explore what minimum ingredients in a
reaction-diffusion system are sufficient to require field
renormalization.

For the special case of $p=1$ in dimension $d=1$ we have determined
$\phi$ from the four walker bracket problem.  This approach strongly
suggests that the exponent should depend on $\delta$, since varying
the parameter from unity has the effect of opening or closing the
angle of the absorbing wedge in the four walker problem, which should
modify the survival probability decay exponent.  This could be
investigated numerically.

We are currently undertaking a simulation of this reaction-diffusion
system that utilizes a variation of the approach of Mehra and
Grassberger \cite{Mehra02}.  These authors studied the trapping
reaction, $A+B\to B$, and developed a method for tracking the entire
$A$ particle distribution conditioned on a realized trajectory of a
single $B$ particle.  For our system, Eq.~(\ref{eq:reaction}), this
can be inverted: the $A$ particle dynamics can be treated via Monte
Carlo, and for a given realization of the $A$ particles, the entire
$B$ particle distribution can be generated.  This method should
allow for reasonably high quality statistics to test the predicted
anomalous scaling for the $C_{BB}$ correlation function, and to
explore the dependence of the dynamical exponents $\theta$ and
$\phi$ on the parameters $p$ and $\delta$.

Finally, the lack of rapid convergence of the $\epsilon$ expansion
appears to be a general feature of these RG fixed points, also
observed in the single-species annihilation reaction \cite{Lee94}.
For that system, Vernon showed that replacing the short-range
diffusive hops with L\'evy flights, governed by a size distribution
$P(r)\sim r^{-d-\sigma}$ with $1 < \sigma < 2$, lowered the upper
critical dimension to $d_c=\sigma$ \cite{Vernon03}.  Thus $\sigma$ can
be chosen so that $\epsilon=\sigma-d$ is small in $d=1$ simulations,
which allowed Vernon to confirm the accuracy of the RG $\epsilon$
expansion \cite{Vernon03}.  Such an approach could be interesting
here, in particular to test whether the $\delta$ dependence of $\phi$
weakens as $\epsilon$ becomes small.

\acknowledgments

J.H. and R.S.M. were supported by NSF REU Grant PHY-1156964 and
J.H. was supported by NSF Grant
DMS-1612921. B.P.V.-L. acknowledges the hospitality of the University
of G\"ottingen, where this work was completed.

\appendix

\section{One Loop Correlation Diagrams}

\begin{figure}
  \includegraphics[width=3.2in]{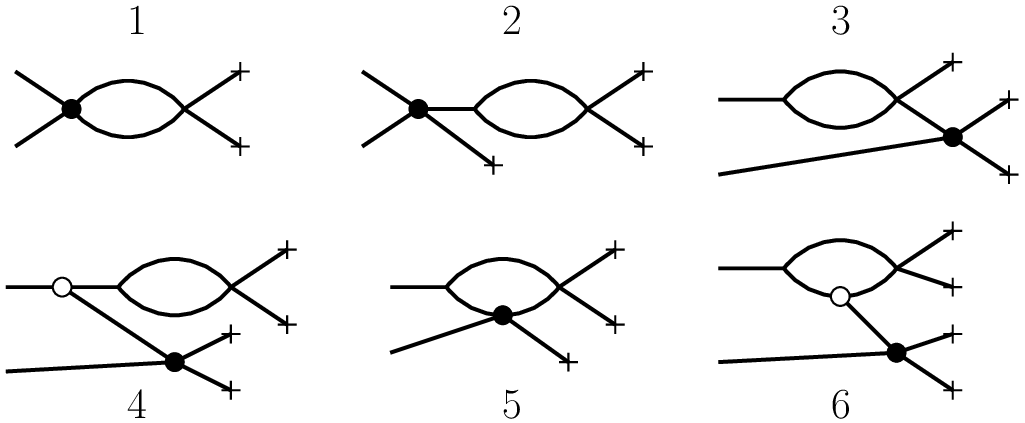}
  \caption{The six topology classes of diagrams (not yet including the
    dashed $B$ particle lines).  The solid circle identifies
    $\lambda_4$ and the open circle $\lambda_3$.  Description:
    (1)~$\lambda_4$ closes loop, (2)~loop in causal past of
    $\lambda_4$, (3)~$\lambda_4$ in causal past of loop, (4)~loop and
    $\lambda_4$ in causal past of $\lambda_3$, (5)~$\lambda_4$ in
    loop, and (6)~$\lambda_3$ in loop.}
  \label{fig:classes}
\end{figure}

Many diagrams contribute to $\hat C_{BB}(k=0)$ at one loop, and care
must be taken to identify them.  All have exactly two four-point
vertices: one that begins the loop and one that links the two terminal
lines.  The latter vertex we label $\lambda_4$.  Some diagrams have a
particular three point vertex, which we label $\lambda_3$, that
connects $\lambda_4$ to the loop.  All possible diagrams fall into six
topology classes, as shown in Fig.~\ref{fig:classes}.  Adding the
dashed $B$ lines to these in every distinct way results in a total of
61 diagrams.  We present our results in the form
\begin{equation}
  \hat C_{BB}^\text{1-loop}(k=0) = \frac{b_0^2 t^{1+\epsilon/2}}{(a_0\lambda
    t)^{2Q\lambda'/\lambda}} \frac{Q\lambda'^2}{12\pi} \sum_{i=1}^6 F_i(z,\epsilon)
\end{equation}
where the $F_i$ are the contributions from each class of diagram,
and $z = \lambda'/\lambda$.

A remark on the order of $\epsilon$: some diagrams contain order
$1/\epsilon^2$ contributions.  As with the density calculation, we
will show that these terms cancel as the couplings flow to their fixed
point values, and $z\to z^*$.  The $1/\epsilon$ portions of the
diagrams will contribute to the renormalization factor $Z_{b^2}$ and
ultimately provide the $O(\epsilon)$ contribution to the anomalous
dimension.  Any diagrams that are finite as $\epsilon\to 0$ do not
contribute to the anomalous dimension and may be neglected.

\begin{figure}
  \includegraphics[width=3.2in]{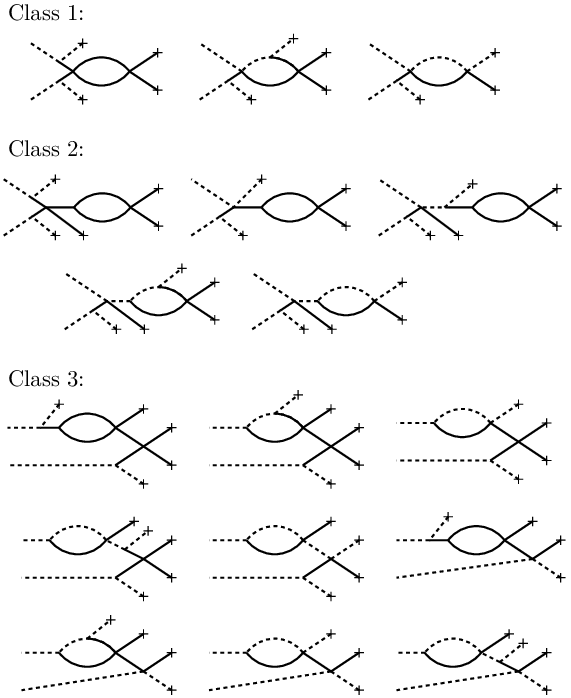}
  \caption{All one-loop diagrams contributing to $\hat C_{BB}(k=0)$ in
  Classes 1, 2, and 3.}
  \label{fig:classes123}
\end{figure}

\begin{figure}
  \includegraphics[width=3.2in]{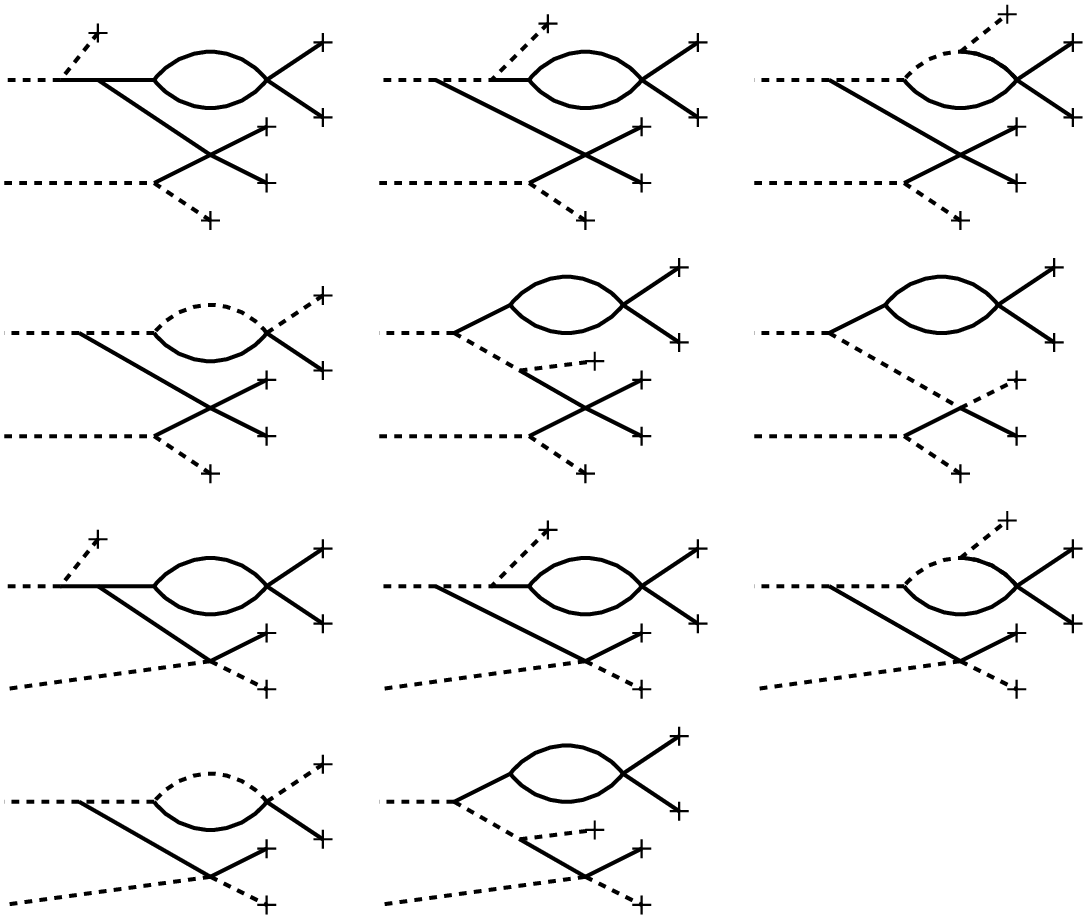}
  \caption{All one-loop diagrams contributing to $\hat C_{BB}(k=0)$ in
  Class 4.}
  \label{fig:class4}
\end{figure}

From the class 1, 2, and 3 diagrams in Fig.~\ref{fig:classes123} we obtain
\begin{align}
  F_1 &=  \biggl( 4Q - \frac{6z}{1+\delta}\biggr)\frac{1}{\epsilon},
  \label{eq:class1}
   \\[1ex]
  F_2 &=
  12 Qz(8\pi)^{\epsilon/2}\biggl(1-\frac{z}{z^*}\biggr)
  \frac{1}{\epsilon^2}  \nonumber\\
  &+\biggl(6+24Qz-4Q+3Q^2z^2f(\delta)-\frac{15 Qz^2}{1+\delta}
  \biggr)\frac{1}{\epsilon},
  \label{eq:class2}
  \\[1ex]
  F_3 &= \biggl(6 - \frac{14Q}{3} + \frac{8Qz-6z-10Q^2z^2+15Qz^2}{1+\delta}
  \biggr)\frac{1}{\epsilon}
  \label{eq:class3}
\end{align}
From the class 4 diagrams in Fig.~\ref{fig:class4} we obtain 
\begin{align}
  F_4 &= \biggl(\frac{4Q}{3}-1\biggr)Q z(8\pi)^{\epsilon/2}
  \biggl(1-\frac{z}{z^*}\biggr)\frac{1}{\epsilon^2} \nonumber\\
  &+\biggl(-6+6Qz+\frac{28Q}{3}-20Q^2z+\frac{18Q^2z^2-12Qz^2}{1+\delta}
  \nonumber\\
  &\qquad +(3Q^2z^2-4Q^3z^2)f(\delta)
  \biggr)\frac{1}{\epsilon}
  \label{eq:class4}
\end{align}

The 18 diagrams in class 5 and the 15 diagrams in class 6 are all
finite as $\epsilon\to 0$ and do not contribute to the anomalous
dimension.  We do not depict these diagrams here.  It is relatively
straightforward to evaluate the order of the diagrams: a factor of
$1/\epsilon$ is obtained from any simple loop that does not have a
propagator attached to it.  No such loops exist in classes 5 and 6.
Some diagrams obtain a factor of $1/\epsilon$ from having a non-loop
time integral of the form $\int t^{-1+\epsilon/2}dt$. These may be
identified by dimensional analysis and do not occur in classes 5 and
6.

\providecommand{\newblock}{}






\newpage


\onecolumngrid

\begin{center}
\textbf{\large Corrigendum: Anomalous Dimension in a Two-Species Reaction-Diffusion
System}

\medskip

Benjamin Vollmayr-Lee$^1$, Jack Hanson$^2$, R. Scott
  McIsaac$^3$, and Joshua D. Hellerick$^1$

  \textit{\small
$^1$Department of Physics \&\ Astronomy, Bucknell University,
  Lewisburg, PA 17837, USA \\
$^2$Department of Mathematics, City College of New York, 160
  Convent Ave, New York, NY 10031 USA \\
$^3$Calico Life Sciences, South San Francisco, CA 94080 USA}

\medskip

\begin{quoting}[leftmargin=0.6in] \small
  We make two corrections to the renormalization group calculation
  presented in J. Phys. A: Math. Theor. 51, 034002 (2018).  First, the
  field renormalization technique presented is not applicable for the
  $B$ particle density in $d=2$ because of noncommutitivity of the
  $\epsilon\to 0$ and $t\to\infty$ limits.  The $B$ particle density
  in $d<2$ and the correlation function for $d\leq 2$ are unaffected
  by this issue.  Second, we correct a symmetry factor in one of the
  diagrams, which modifies the correlation function scaling exponents.
\end{quoting}


\medskip

\end{center}

\twocolumngrid

The renormalization group calculation presented in
\cite{VollmayrLee2018} contains two errors.  The first concerns the
$B$ particle density at the upper critical dimension $d=2$, which
decays as $\langle b\rangle \sim t^{-\theta}(\ln t)^\alpha$.  Our
calculation of $\alpha$ was based on the assumption that the
contribution $A(z)/\epsilon^2$ in Eq.~(25) was asymptotically
negligible because
\begin{equation} \tag{1}
  A(z)\propto (z-z^*) \sim \begin{cases} t^{-\epsilon/2} & d<2 \\
    1/\ln t & d=2.
    \end{cases}
\end{equation}
This assumption is valid for $d<2$ since for any $\epsilon>0$, there
exists an asymptotic regime where $1/(\epsilon^2 t^{\epsilon/2})$ is
negligibly small.  But for $d=2$ we must take the $\epsilon\to 0$
limit before the large $t$ limit, and this term cannot be neglected.
As a result, the field renormalization technique employed in
\cite{VollmayrLee2018} is not applicable and one must instead employ
the technique of Rajesh and Zaboronski \cite{Rajesh2004}, where they
renormalize instead the logarithmic derivative $t\partial_t \ln
\langle b\rangle$.  Their result agrees with our Ref.~[1] Eq.~(7), but their
value of $\alpha$, which in our notation reads
\begin{align}
  \alpha =& \biggl(\frac{1+\delta}{2-p}\biggr)
  \biggl[\frac{3}{2}+\ln\biggl(\frac{1+\delta}{2}\biggr)+
    \frac{1}{2}\biggl( \frac{1+\delta}{2-p}\biggr) f(\delta)\biggr]
  \nonumber\\ & - \frac{4\pi(1+\delta)}{2-p}\biggl(\frac{1}{\lambda} -
  \frac{1+\delta}{\lambda'}\biggr),  \tag{2}
  \label{eq:alpha}
\end{align}
corrects the value we reported in Eq.~(8). 

However, the problem of the noncommuting $\epsilon\to 0$ and
$t\to\infty$ limits does not affect the calculation of correlation
function scaling exponents $\phi$ and $\alpha_2$, defined via
\begin{equation} \tag{3}
  \tilde C_{BB}(r,t) = \begin{cases} t^\phi f(r/\sqrt t) & d<2 \\
    (\ln t)^{\alpha_2} f(r/\sqrt t) & d=2, \end{cases}
\end{equation}
provided one renormalizes the \textit{scaled} correlation function
$\tilde C_{BB}(r,t) = C_{BB}(r,t)/\langle b(t)\rangle^2$.  Our bare
expansion for the unscaled $C_{BB}^B$ in Eq.~(40) is then replaced by
\begin{align}
  \hat{\tilde C}_{BB}^B = \frac{t\lambda'^2 h(Q)}{\lambda}
  \biggl[ 1&+\lambda t^{\epsilon/2}
    \biggl( \frac{C(z)-2A(z)}{\epsilon^2}\nonumber\\
    &+ \frac{D(z)-2B(z)}{\epsilon} +\dots  \tag{4}
    \biggr)\biggr].
\end{align}
The $1/\epsilon^2$ term then vanishes because $C(z)=2A(z)$.  The
remaining $1/\epsilon$ term can be controlled by field renormalization
as before, with the same final results:
\begin{equation}  \tag{5}
  \phi = \biggl[ \pi \Bigl(D(z^*) - 2 B(z*)\Bigr) + \frac{1}{2}\biggr]\epsilon +
  O(\epsilon^2)
\end{equation}
for $d<2$ and
\begin{equation} \tag{6}
  \alpha_2 =2\pi \Bigl(D(z^*)-2B(z^*)\Bigr)
\end{equation}
for $d=2$.

The second error in \cite{VollmayrLee2018} was a symmetry factor of two
in the first diagram of Class 2 in Fig.~A2.  The corrected Eq.~(A.3) reads
\begin{align}
  F_2 =& 12Qz(8\pi)^{\epsilon/2}\biggl(\frac{z}{z^*}-1\biggr)\frac{1}{\epsilon^2}
  \nonumber\\
  &+ \Biggl(6+24Qz-8Q+3Q^2z^2f(\delta)-\frac{15Qz^2}{1+\delta}\biggr)  \tag{7}
  \frac{1}{\epsilon}
\end{align}
which changes $D(z^*)$ in Eq.~(41) to
\begin{equation}
  D(z^*) = -\frac{9-13Q}{6\pi(3-2Q)} + \frac{3Q(1+\delta)}{2\pi}
  +\frac{Q^2(1+\delta)^2f(\delta)}{2\pi} + O(\epsilon).  \tag{8}
\end{equation}
This modifies the correlation function exponents:
Eq.~(6) becomes
\begin{equation} \tag{9}
  \phi = \frac{7}{24-18p}\epsilon + O(\epsilon^2)
\end{equation}
and Eq.~(10) becomes
\begin{equation} \tag{10}
  \alpha_2 = - \frac{5-9p}{12-9p}
\end{equation}
In the text after Eq.~(11), the value of $\phi$ for the truncated
RG expansion in $d=1$ with $p=1$ and $\delta=1$ is $\phi = \frac{7}{6}
\simeq 1.17$.


\providecommand{\newblock}{}

\end{document}